\documentclass{sf2a-conf2011}
\usepackage{graphicx}
\usepackage{hyperref}
\usepackage[]{natbib}  
\usepackage[cyr]{aeguill}
\usepackage{epstopdf}

\def\BibTeX{{\rm B\kern-.05em{\sc i\kern-.025em b}\kern-.08em
    T\kern-.1667em\lower.7ex\hbox{E}\kern-.125emX}}
\bibpunct{(}{)}{;}{a}{}{,}  %%%%%%%%%%%%%  A&A bibliography style
\begin{document}

\TitreGlobal{SF2A 2011}

%%-----------------------------------------------------------------
%%      the top matter
%%

\title{The Dependence of the Galactic Star Formation Laws on Metallicity}

\runningtitle{The dependence of star formation rates in galaxies on metallicity}

\author{S. Dib}\address{Astrophysics Group, Blackett Laboratory, Imperial College London, London SW7 2AZ; s.dib@imperial.ac.uk}
\author{L. Piau}\address{LATMOS, 11 Boulevard d'Alembert, 78280 Guyancourt, France}
\author{S. Mohanty$^{1}$}
\author{J. Braine}\address{Laboratoire d'Astrophysique de Bordeaux, Universit\'{e} de Bordeaux, OASU CNRS/INSU, 33271 Floirac, France}

%% Keep this line, even if the page will be settled afterwards.
\setcounter{page}{237}

%% To make the final index, repeat the authors here, in the format : Surname, Initial(s) 
\index{Dib, S.}
\index{Piau, L}
\index{Mohanty, S.}
\index{Braine, J.}

%%-----------------------------------------------------------------

\maketitle

%%-----------------------------------------------------------------
%%        The abstract
%% 
%%  Warning!  within the abstract:
%%  - do not use macros. 
%%  - do not use commands like: \cite, \citet, \citep ... etc.

\begin{abstract}
We describe results from semi-analytical modelling of star formation in protocluster clumps of different metallicities. In this model, gravitationally bound cores form uniformly in the clump following a prescribed core formation efficiency per unit  time. After a contraction timescale which is equal to a few times their free-fall times, the cores collapse into stars and populate the IMF. Feedback from the newly formed OB stars is taken into account in the form of stellar winds. When the ratio of the effective energy of the winds to the gravitational energy of the system reaches unity, gas is removed from the clump and core and star formation are quenched. The power of the radiation driven winds has a strong dependence on metallicity and it increases with increasing metallicity. Thus, winds from stars in the high metallicity models lead to a rapid evacuation of the gas from the protocluster clump and to a reduced star formation efficiency, as compared to their low metallicity counterparts. We derive the metallicity dependent star formation efficiency per unit time in this model as a function of the gas surface density $\Sigma_{g}$. This is combined with the molecular gas fraction in order to derive the dependence of the surface density of star formation $\Sigma_{SFR}$ on $\Sigma_{g}$. This feedback regulated model of star formation reproduces very well the observed star formation laws in galaxies extending from low gas surface densities up to the starburst regime. Furthermore, the results show a dependence of $\Sigma_{SFR}$ on metallicity over the entire range of gas surface densities, and can also explain part of the scatter in the observations.
\end{abstract}


%%-----------------------------------------------------------------

\section{Introduction}

Over the last two decades, the dependence of the star formation rate surface density ($\Sigma_{SFR}$) on the gas surface density ($\Sigma_{g}$) and eventually on other physical quantities has been extensively investigated both observationally (e.g., Kennicutt 1998; Wong \& Blitz 2002; Boissier et al. 2003; Bigiel et al. 2008; Blanc et al. 2009; Onodera et al. 2010; Tabatabaei \& Berkhuijsen 2010; Heiner et al. 2010; Schruba et al. 2011; Bolatto et al. 2011) as well as theoretically and numerically (e.g., Tutukov 2006; Krumholz \& Thompson 2007; Fuchs et al. 2009; Silk \& Norman 2009; Krumholz et al. 2009a; Papadopoulos \& Pelupessy 2010; Gnedin \& Kravtsov 2011; Narayanan et al. 2011; Feldmann et al. 2011; Vollmer \& Leroy 2011; Braun \& Schmidt 2011; Monaco et al. 2011; Kim el al. 2011; Dib 2011a,b). Determining the rate of star formation in a given tracer of the gas surface density requires quantifying the fraction of that tracer as a function of the global gas surface density and a description of the efficiency at which the star forming gas is converted into stars per unit time. The relationship between $\Sigma_{SFR}$ and the surface density of the molecular hydrogen gas $\Sigma_{H_{2}}$ is given by:                
  
\begin{equation} 
\Sigma_{SFR}=\Sigma_{g}~f_{H_{2}} \frac{SFE_{\tau}}{\tau},
\label{eq_1_1}
\end{equation} 

\noindent where $f_{H_{2}}=\Sigma_{H_{2}}/\Sigma_{g}$ is the molecular hydrogen mass fraction is the atomic-molecular star forming complexes, and $SFE_{\tau}$ is the star formation efficiency over the timescale $\tau$. Krumholz \& McKee (2005) proposed a theory in which supersonic turbulence is the dominant agent that regulates star formation in giant molecular clouds (GMCs). They derived a core formation efficiency per unit free-fall time, $CFE_{ff}$, which is given by $CFE_{ff}=0.15 \alpha_{vir}^{-0.68} {\cal M}^{-0.32}$, where $\alpha_{vir}$ and ${\cal M}$ are the virial parameter and the $\it{rms}$ sonic Mach number of the GMC, respectively\footnote{Padoan \& Nordlund (2011) found a different dependence of the SFR on $\alpha_{vir}$ and ${\cal M}$. The results of their numerical simulations suggest that the SFR decreases with increasing $\alpha_{vir}$ but also that it increases with increasing ${\cal M}$.}. By assuming that only a fraction of the mass of the cores ends up in the stars, this $CFE_{ff}$ can be converted into a star formation efficiency $SFE_{ff}=\eta \times CFE_{ff}$ ($\eta \leq 1$). An alternative theory has been recently proposed by Dib et al. (2011) and Dib (2011a,b) in which the star formation rate in protocluster clumps is primarily regulated by feedback from massive stars and in particular through energy injection in the clumps by stellar winds. In the following sections, we briefly describe the main constituents of this model. 
  
\section{Feedback regulated star formation}

The model follows the formation of dense gravitationally bound cores in a protocluster clump. Cores form in the clump with a given core formation efficiency per unit time and follow a local mass distribution that is the result of the gravo-turbulent fragmentation of the clump. In their series of models, Dib et al. (2011) varied the core formation efficiency per unit free-fall time ($CFE_{ff}$) between $0.1$ and $0.3$. This is consistent with the range of $CFE_{ff}$ measured in numerical simulations which describe the gravo-turbulent fragmentation of magnetised, turbulent, and self-gravitating molecular clouds (e.g., Dib et al. 2008; Dib et al. 2010a). The gravitationally bound cores that are generated at every epoch have a mass distribution that is given by the gravo-turbulent fragmentation model of Padoan \& Nordlund (2002). In this work, we leave out, for simplicity, the role played by gas accretion and coalescence in modifying the mass distribution of the cores. The interested reader is referred to Dib et al. (2007) and Dib et al. (2010b) for such models. Cores contract over a lifetime which is a few times their free-fall time before collapsing to form stars. Feedback from the most massive stars ($M_{\star} \geq 5$ M$_{\odot}$) is taken into account in the form of stellar winds. The formation of cores in the protocluster clump, and consequently star formation, are terminated whenever the fraction of the wind energy stored into motions that oppose gravity exceeds the gravitational energy of the clump.  In order to calculate reliable estimates of the feedback generated by metallicity dependent stellar winds, we proceed in two steps. In the first step, we use a modified version of the stellar evolution code CESAM (see appendix 1 in Piau et al. 2011) to calculate a grid of main sequence stellar models for stars in the mass range [5-80] M$_\odot$ (with steps of 5 M$_{\odot}$) at various metallicities $Z/Z_{\odot}=[1/10, 1/6, 1/3, 1/2, 1, 2]$ ($Z_{\odot}=0.0138$). The evolution of massive stars is followed using the CESAM code for $\sim 1$ Myr, on the main sequence. The characteristic stellar properties, which are the effective temperature $T_{eff}$, the luminosity $L_{\star}$, and the stellar radius $R_{\star}$ are then used in the stellar atmosphere model of Vink et al. (2001) in order to calculate the stellar mass loss rate $\dot{M}_{\star}$. Vink et al. (2001) did not derive the values of the terminal velocities of the winds ($v_{\infty}$), therefore, we use instead the derivations of $v_{\infty}$ obtained by Leitherer et al. (1992). 

\begin{figure}[ht!]
\centering
\includegraphics[width=0.8\textwidth]{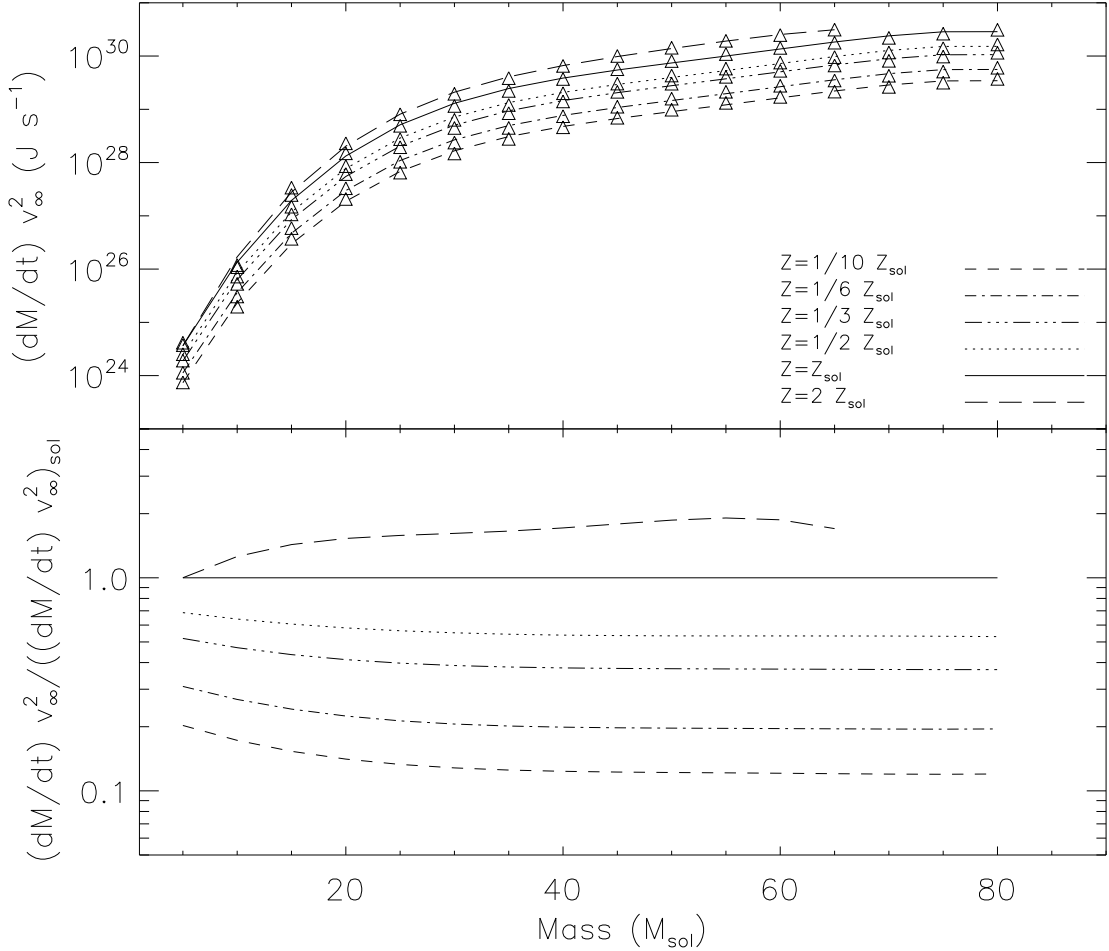}      
\vspace{1cm}
\caption{ The power of the stellar winds, or wind luminosities, for stars in the mass range 5-80 M$_{\odot}$ on the main sequence, and for various metallicities. The stellar mass loss rates have been calculated using the stellar characteristics (effective temperature, stellar luminosity and radius) computed using the stellar evolution code CESAM coupled to the stellar atmosphere model of Vink et al. (2001). The values of $v_{\infty}$ have been calculated using the derivation by Leitherer et al. (1992). Over-plotted to the data are fourth order polynomials. The parameters of the fit functions can be found in Dib et al. (2011). Adapted from Dib et al. (2011).}
 \label{dib_f1}
\end{figure}

The power of the stellar winds is given by $\dot{M}_{\star} v_{\infty}^{2}$. This quantity is displayed in Fig.~\ref{dib_f1} for the models with different metallicities. The values of $\dot{M}_{\star} v_{\infty}^{2}$ are fitted with fourth order polynomials (overplotted to the data) and whose coefficients are provided in Dib et al. (2011). The $\dot{M}_{\star} v_{\infty}^{2}-M_{\star}$ relations displayed in Fig.~\ref{dib_f1} allow for the calculation of the total wind energy deposited by stellar winds. The total energy from the winds is given by:

\begin{equation}
E_{wind} = \int_{t'=0}^{t'=t} \int_{M_{\star}=5~\rm{M_{\odot}}}^{M_{\star}=120~\rm{M_{\odot}}} \left( \frac{N(M_{\star}) \dot{M_{\star}} (M_{\star}) v_{\infty}^{2}}{2} dM_{\star}\right) dt'. 
\label{eq_2_1}
\end{equation}

We assume that only a fraction of $E_{wind}$ will be transformed into systemic motions that will oppose gravity and participate in the evacuation of the bulk of the gas from the proto-cluster clump. The effective kinetic wind energy is thus given by: 
 
\begin{equation}
E_{k,wind}=\kappa~E_{wind},
\label{eq_2_2}
\end{equation}

\noindent where $\kappa$ is a quantity $\leq 1$ (in this work, we use $\kappa=0.1$ for all models). $E_{k,wind}$ is compared at every timestep to the absolute value of the gravitational energy, $E_{grav}$, which is calculated as being:

\begin{equation}
E_{grav} =  -\frac{16}{3} \pi^{2} G \int_{0}^{R_{c}} \rho_{c}(r)^{2} r^{4}  dr.
\label{eq_2_3}
\end{equation}

\begin{figure}[ht]
\centering
\includegraphics[width=0.8\textwidth]{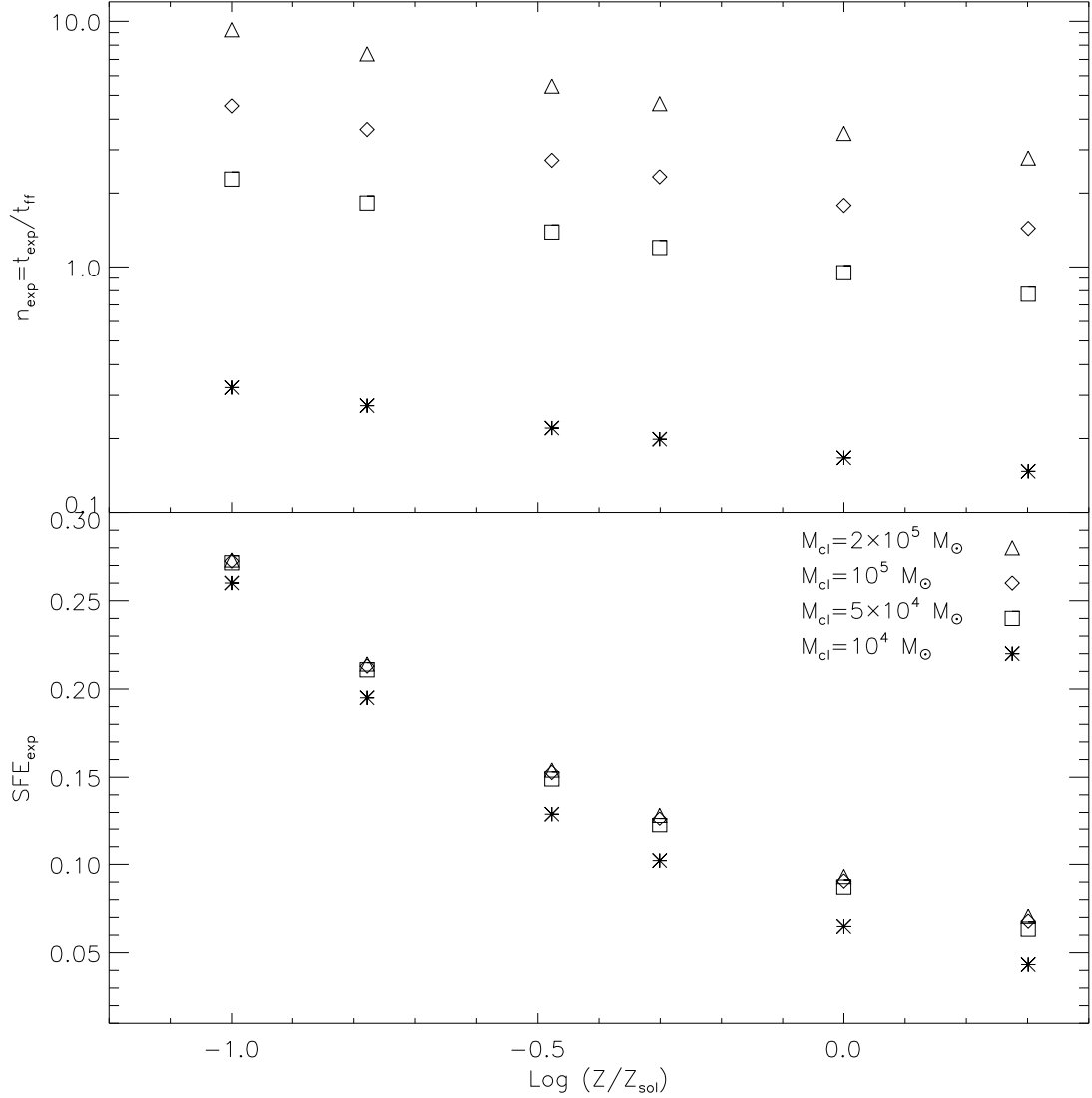} 
\vspace{1cm} 
\caption{Dependence of the quantities $SFE_{exp}$ (final star formation efficiency) and $n_{exp}=t_{exp/t_{ff}}$ (ratio of the expulsion time to the free-fall time) for selected values of the protocluster forming clump masses and metallicities. These results are based on the models of Dib et al. (2011).}
\label{dib_f2}
\end{figure}

Since higher metallicity stellar winds deposit larger amounts of energy in the clump than their lower metallicity counterparts, this leads them to evacuate the gas from the clump on shorter timescales. This in turn quenches the process of core and star formation earlier and sets a smaller final star formation efficiency, $SFE_{exp}$. Fig.~\ref{dib_f2} displays the dependence of $SFE_{exp}$ and of the expulsion time, $t_{exp}$ (expressed in units of the free-fall time $t_{ff}$), as a function of metallicity for clumps of various masses. 

\section{Metallicity dependent star formation laws}

Using the above described model, it is possible to derive the dependence of $\Sigma_{SFR}$ on $\Sigma_{g}$. The star formation rate surface density in the feedback regulated mode of star formation is given by:

\begin{equation} 
\Sigma_{SFR}=\Sigma_{g}~f_{H2} \frac{\left<SFE_{exp}\right>} {\left<t_{exp}\right>},
\label{eq_3_1}
\end{equation}  

\noindent where $\left<SFE_{exp}\right>$ and $\left<t_{exp}\right>$ are, respectively, the characteristic $SFE_{exp}$ and the epoch at which gas is expelled from the protocluster region for the clump mass distribution associated with a given $\Sigma_{g}$. Writing $\left<t_{exp}\right>$ in terms of the clump free-fall time $\left<t_{ff}\right>$, Eq.~\ref{eq_3_1} becomes:

\begin{eqnarray}
\Sigma_{SFR}=\Sigma_{g}~f_{H_{2}} \frac{\left<SFE_{exp}\right>} {\left<n_{exp}\right>} \frac{1}{\left<t_{ff}\right>} \nonumber \\
                                 ~=\Sigma_{g}~f_{H_{2}} \frac{\left<f_{\star,ff}\right>}{\left<t_{ff}\right>}.
\label{eq_3_2}                  
\end{eqnarray}

\noindent where $f_{\star,ff}$ is the dimensionless star formation efficiency and which corresponds to the mass fraction of the molecular gas that is converted into stars per free-fall time $t_{ff}$ of the clumps. $\left< f_{\star,ff}\right>$ and $\left< t_{ff}\right>$ represent characteristic values of $f_{\star,ff}$ and $t_{ff}$ for the spectrum of clump masses found in the GMC for a given value of $\Sigma_{g}$. The quantity $f_{H_{2}}$ is the mass fraction of the total gas that is in molecular form. In this work, we use the functional form of $f_{H_{2}}$ obtained by Krumholz et al. (2009b) who derived $f_{H_{2}}$ as a function of the gas surface density and metallicity (see their paper or Dib 2011 for the detailed formula). $\left<t_{ff}\right>$ can be approximated by the free-fall time of the clump with the characteristic mass $t_{ff} (M_{char})=8 \Sigma_{cl}^{' -3/4} M_{char,6}^{1/4}$ Myr where $M_{char,6}=M_{char}/10^{6}$ M$_{\odot}$. The characteristic mass $M_{char}$ is given by :

\begin{figure}[ht]
\centering
\includegraphics[width=0.8\textwidth]{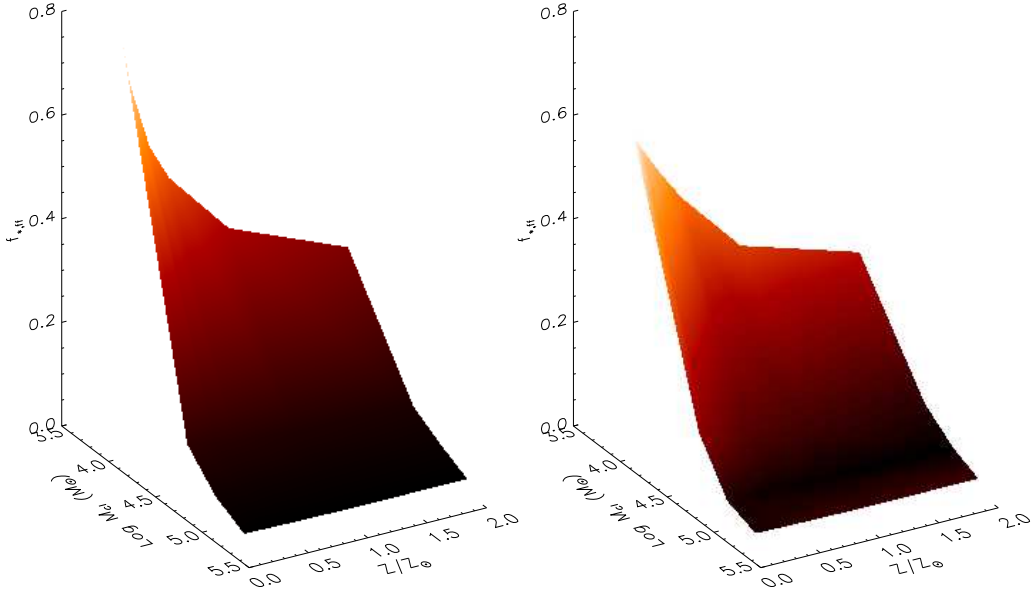} 
\caption{\label{fig11} {\small Star formation efficiency per unit free-fall time in the protocluster clump in the metallicity-dependent feedback model. The star formation efficiencies per free-fell use a core-to-star efficiency conversion factor of 1/3. The left panel displays $f_{\star,ff}$ as a function of both $M_{cl}$ and $Z^{'}=Z/Z_{\odot}$ in the original data, while the right panel displays the analytical fit function to this data set given in Eq.~\ref{eq_3_5}. Adapted from Dib (2011a).}
}
\label{dib_f3}
\end{figure}

\begin{equation}
 M_{char}= \int_{M_{cl,min}}^{max(M_{cl,max},M_{GMC})} M_{cl} N(M_{cl}) dM_{cl}, 
 \label{eq_3_3}
\end{equation}

\begin{figure}[ht]
\centering
\includegraphics[width=0.8\textwidth]{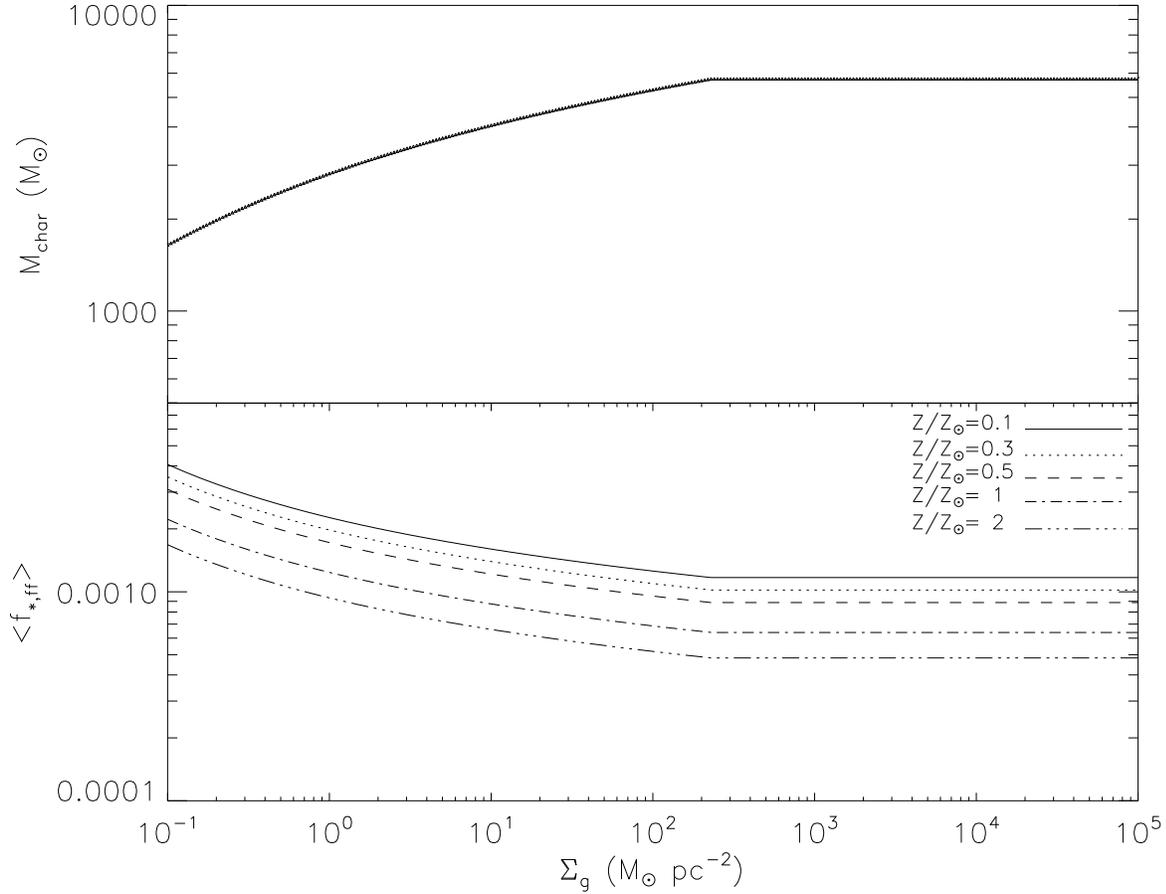} 
\vspace{1cm}
\caption{\label{fig12} {\small Characteristic clump mass as a function of the gas surface density (Eq.~\ref{eq_3_3}, top panel) and the star formation efficiency per unit free-fall time in this feedback regulated model of star formation (lower panel). Adapted from Dib (2011a).}
}
\label{dib_f4}
\end{figure}

\noindent where $N (M_{cl})$ is the mass function of protocluster forming clumps which we take to be $N (M_{cl})=A_{cl} M_{cl}^{-2}$, and $A_{cl}$ is a normalisation coefficient given by $A_{cl} \int_{M_{cl,min}}^{max(M_{cl,max},M_{GMC})} N(M_{cl}) dM_{cl}= \epsilon$, where $0 < \epsilon < 1$ is the mass fraction of the GMCs that is in protocluster clumps at any given time. In this work we use $\epsilon=0.5$. The minimum clump mass $M_{cl,min}$ is taken to be $2.5 \times 10^{3}$ M$_{\odot}$ (this guarantees, for final SFEs  in the range of 0.05-0.3 a minimum mass for the stellar cluster of $\sim 50$ M$_{\odot}$) and the maximum clump mass is $10^{8}$ M$_{\odot}$. The characteristic GMC mass is determined by the local Jeans mass and is given by:

\begin{equation} 
M_{GMC}=37 \times 10^{6} \left(\frac{\Sigma_{g}}  {85~\rm{M_{\odot}~pc^{-2}}} \right) {\rm M_{\odot}}. 
\label{eq_3_4}
\end{equation} 

Fig.~\ref{dib_f4} (top) displays $M_{char}$ as a function of $\Sigma_{g}$. The quantity $f_{\star,ff}=SFE_{exp}/n_{exp}$ is displayed in Fig.~\ref{dib_f3} (left panel) as a function of mass and metallicity ($Z^{'}=Z/Z_{\odot}$). These models use a value of $CFE_{ff}=0.2$ and standard clump and core parameters (see Dib et al. 2001 and Dib 2011a,b for more detail). A fit to the ($M_{cl},Z^{'}$) data points with a 2-variables second order polynomial yields the following relation shown in Fig.~\ref{dib_f3}, right panel:

\begin{eqnarray}
f_{\star,ff}(M_{cl},Z^{'})=  11.31-4.31{\rm log(M_{cl})} + 0.41 {\rm [log(M_{cl})]^{2}}  \nonumber \\ 
           - 8.28 Z^{'} + 3.20 Z^{'} {\rm log (M_{cl})} - 0.32 Z^{'} {\rm [log(M_{cl})]^{2}}  \nonumber \\
          + 2.30 Z^{'2} - 0.89 Z^{'2} {\rm log(M_{cl})} + 0.08 Z^{'2} {\rm [log(M_{cl})]^{2}}.
\label{eq_3_5}
\end{eqnarray} 

\begin{figure}[ht]
\centering
 \includegraphics[width=0.8\textwidth]{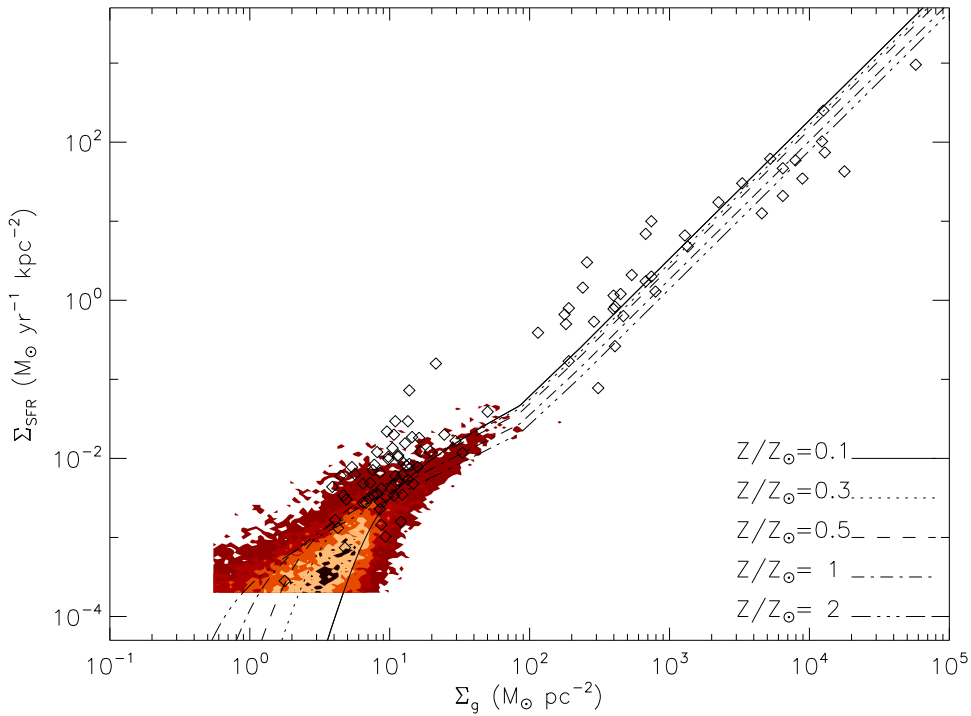}     
\caption{Star formation laws in the feedback-regulated star formation model. Overplotted to the models are the normal and starburst galaxies data of Kennicutt (1998) and the combined sub-kpc data (4478 subregions) for 11 nearby galaxies from Bigiel et al. (2008,2010). The Bigiel et al. data is shown in the form of a 2D histogram with the colour coding corresponding, from the lighter to the darker colours to the 1,5,10,20, and 30 contour levels. The displayed theoretical models cover the metallicity range $Z^{'}=Z/Z_{\odot}=[0.1,2]$.}
 \label{dib_f5}
\end{figure}

Using Eq.~\ref{eq_3_5}, it is then possible to calculate $\left<f_{\star,ff}\right>$: 

\begin{equation}
\left<f_{\star,ff}\right> (Z^{'}, \Sigma_{g}) = \int_{M_{cl,min}}^{max(M_{cl,max},M_{GMC})} f_{\star,ff} (M_{cl},Z^{'}) N(M_{cl}) dM_{cl}. 
\label{eq_3_6}
\end{equation}

Fig.~\ref{dib_f4} (bottom) displays $\left<f_{\star,ff}\right>$ ($Z^{'},\Sigma_{g}$) for values of $Z^{'}$ in the range [$0.1-2$]. We assume that there is a critical value of $\Sigma_{g}= 85$ M$_{\odot}$ pc$^{-2}$ below which clumps are pressurised by their internal stellar feedback, such that $\Sigma_{cl}=\Sigma_{g,crit}$ where $\Sigma_{g} < \Sigma_{g,crit}$ and $\Sigma_{cl}=\Sigma_{GMC}=\Sigma_{g}$ when $\Sigma_{g} \geq \Sigma_{g,crit}$. With the above elements, the star formation law can be re-written as:

\begin{eqnarray}
\Sigma_{SFR} &=& \frac{8} {10^{6}} f_{H_{2}} (\Sigma_{g},c,Z^{'}) \Sigma_{g}  \nonumber \\
  \times & & \left \{\begin{array} {cc}  \frac{\left<f_{\star,ff}\right> (Z^{'})} {M_{char,6}^{1/4}} &  ; \frac{\Sigma_{g}} {85~{\rm M_{\odot}~pc^{-2}}} < 1\\
  \frac{\left<f_{\star,ff}\right> (Z^{'})} {M_{char,6}^{1/4}} \left(\frac{\Sigma_{g}}{85~{\rm M_{\odot} pc^{-2}}}\right)^{3/4}  & ; \frac{\Sigma_{g}}{85~{\rm M_{\odot}~pc^{-2}}} \geq 1 \end{array} \right \},
\label{eq_3_7}
\end{eqnarray}

\noindent where $\Sigma_{SFR}$ is in M$_{\odot}$ yr$^{-1}$ kpc$^{-2}$, $M_{char}$ is given by Eq.~\ref{eq_3_3}, and $\left<f_{\star,ff}\right>$ by Eqs.~\ref{eq_3_5} and \ref{eq_3_6}. Fig.~\ref{dib_f5}  displays the results obtained using Eq.~\ref{eq_3_7} for $\Sigma_{g}$ values starting from low gas surface densities up to the starburst regime. The results are calculated for the metallicity values of $Z^{'}=[0.1,0.3,0.5,1,2]$. The results are compared to the sub-kpc data of Bigiel et al. (2008,2010) and to the normal and starburst galaxies results of Kennicutt (1998). The models fits remarkably well the observational results over the entire range of surface densities. Furthermore, the segregation by metallicity extends beyond the low surface density regime up to the starburst regime where a segregation in metallicity of $\sim 0.5$ dex is observed. 

\section{Conclusions}
We have presented a model for star formation in protocluster clumps of difference metallicities. The model describes the co-evolution of the dense core mass function and of the IMF in the clumps. Cores form uniformly over time in the clumps following a prescribed core formation efficiency per unit time. Cores contract over timescales which are a few times their free fall time before they collapse to form stars. Feedback from the newly formed OB stars ($> 5$ M$_{\odot}$) is taken into account and when the ratio of the cumulated effective kinetic energy of the winds to the gravitational energy of the system (left over gas+stars) reaches unity, gas is expelled from the clump and further core and star formation are quenched. The radiation driven winds of OB stars are metallicity dependent. Metal rich OB stars inject larger amount of energy into the clump than their low metallictiy counterparts and thus help expel the gas on shorter timescales. This results in reduced final star formation efficiencies in metal rich clumps in comparison to their low metallicity counterparts. Both the final star formation efficiency and the gas expulsion timescales are combined for a grid of clump models with different masses and metallicities in order to calculate the star formation efficiency per unit time ($f_{\star,ff}$) in this feedback regulated model of star formation. We calculate the characteristic value of $f_{\star,ff}$ for a clump mass distribution associated with a gas surface density, $\Sigma_{g}$. This is combined with a description of the molecular mass fraction as a function of $\Sigma_{g}$ and the assumption that there is a critical surface gas density ($\Sigma_{g}=85$ M$_{\odot}$ pc$^{-2}$) above which the protocluster clumps and their parent giant molecular clouds switch from being pressurised from within by stellar feedback to being confined by the external interstellar medium pressure. The combination of these three elements allows us to construct the star formation laws in galaxies going from low gas surface densities up to the starburst regime. Our models exhibit a dependence on metallicity over the entire range of considered gas surface densities and fits remarkably well the observational data of Bigiel et al. (2008,2010) and Kennicutt (1998). This dependence on metallicity of the KS relation may well explain the scatter (or part of it) that is seen in the observationally derived relations. 

\begin{acknowledgements}
I would like to thank the organizers of the workshop {\it Stellar and Interstellar physics for the modelling of the Galaxy and its components} for the opportunity to speak, and would like to acknowledge the generous financial support from the SF2A. S.D. and S.M. acknowledge the support provided by STFC grant ST/H00307X/1.
\end{acknowledgements}

%%-----------------------------
%%   Bibliography
%%-----------------------------

\bibliographystyle{aa}  % A&A bibliography style file (aa.bst)
\bibliography{sf2a-template} % your references in file: Yourfile.bib

\begin{thebibliography}{}

\bibitem[Bigiel et al. (2008)] {bigiel08} Bigiel, F., Leroy, A., Walter, F. et al.  2008, AJ, 136, 2846 
\bibitem[Bigiel et al. (2010)] {bigiiel10} Bigiel, F., Leroy, A., Walter, F. et al.  2010, AJ, 140, 1194
\bibitem[Blanc (2009)] {blanc09} Blanc, G. A., Heiderman, A., Gebhardt, K., Evans, N. J. II, \& Adams, J. 2009, ApJ, 704, 842
\bibitem[Boissier et al. (2003)] {boissier03} Boissier, S., Prantzos, N., Boselli, A., \& Gavazzi, A. 2003, MNRAS, 346, 1215
\bibitem[Bolatto et al. (2011)] {bolatto11} Bolatto, A. D., Leroy, A. K., Jameson, K. et al. 2011, ApJ, accepted, (arXiv:1107.1717)
\bibitem[Boquien (2011)] {boquien11} Boquien, M., Lisenfeld, U., Duc, P.-A. et al. 2011, A\&A, 533, 19
\bibitem[Braun (2011)] {braun11} Braun, H., \& Schmidt, W. 2011, MNRAS, submitted, (arXiv:1104.5582)
\bibitem[Dib (2007)] {dib07} Dib, S., Kim, J., \& Shadmheri, M. 2007, MNRAS, 381, L40 
\bibitem[Dib (2008)] {dib08} Dib, S., Brandenburg, A., Kim, J., Gopinathan, M., \& Andr\'{e}, P. 2008, ApJ, 678, L105
\bibitem[Dib (2010a)] {dib10a} Dib, S., Hennebelle, P., Pineda, J. E., Csengeri, T., Bontemps, S., Audit, E., \& Goodman, A. A. 2010a, ApJ, 723, 425
\bibitem[Dib (2010b)] {dib10b} Dib, S., Shadmehri, M., Padoan, P., Maheswar, G., Ojha, D. K., \& Khajenabi, F. 2010b, MNRAS, 405, 401 
\bibitem[Dib et al. (2011)] {dibetal11} Dib, S., Piau, L., Mohanty, S., \& Braine, J. 2011, MNRAS, 415, 3439
\bibitem[Dib (2011a)] {dib11a} Dib, S. 2011a, ApJ, 737, L20
\bibitem[Dib (2011b)] {dib11b} Dib, S. 2011b, in  Stellar Clusters and Associations- A RIA workshop on GAIA, (arXiv:1107.0886)
\bibitem[Feldmann et al (2011)] {feldmann11} Feldmann, R., Gnedin, N. Y., \& Kravtsov, A. V. 2011, ApJ, 732, 115 
\bibitem[Fuchs (2009)] {fuchs09} Fuchs, B., Jahrrei\ss, H., \& Flynn, C. 2009, 137, 266
\bibitem[Gnedin (2011)] {gnedin11} Gnedin, N. Y., \& Kravtsov, A. V. 2011, ApJ, 728, 88 
\bibitem[Heiner (2010)] {heiner10} Heiner, J. S., Allen, R. J., \& van der Kruit 2010, ApJ, 719, 1244
\bibitem[Kennicutt (1989)] {kennicutt89} Kennicutt, R. C. Jr. 1989, ApJ, 344, 685
\bibitem[Kennicutt (1998)] {kennicutt98} Kennicutt, R. C. Jr. 1998, ApJ, 498, 541
\bibitem[Kim (2011)] {kim11} Kim, C.-G., Kim, W.-T., Ostriker, E. C. 2011, ApJ, accepted, (arXiv:1109.0028) 
\bibitem[Krumholz (2005)] {krumholz05} Krumholz, M. R. \& McKee, C. F. 2005, ApJ, 630, 250 
\bibitem[Krumholz (2007)] {krumholz07} Krumholz, M. R., \& Thompson, T. A. 2007, ApJ, 669, 289
\bibitem[Krumholz (2009a)] {krumholz09a} Krumholz, M. R., McKee, C. F., \& Tumlinson, J. 2009, ApJ, 699, 850
\bibitem[Krumholz (2009b)] {krumholz09b} Krumholz, M. R., McKee, C. F., \& Tumlinson, J. 2009, ApJ, 693, 216
\bibitem[leitherer (1992)] {leitherer92} Leitherer, C., Robert, C., \& Drissen, L. 1992, ApJ, 401, 596
\bibitem[Monaco (2011)] {monaco11} Monaco, P., Murante, G., Borgani, S., \& Dolag, K. 2011, MNRAS, accepted, (arXiv:1109.0484)
\bibitem[Narayanan et al. (2011)] {narayanan11} Narayanan, D., Cox, T. J., Hayward, C. C., Hernquist, L. 2011, MNRAS, 412, 287
\bibitem[Onodera (2011)] {onodera11} Onodera, S., Kuno, N., Tosaki, T. et al. 2010, ApJ, 722, 127
\bibitem[Padoan (2002)] {padoan02} Padoan, P., \& Nordlund, \AA. 2002, ApJ, 576, 870 
\bibitem[Padoan (2011)] {padoan11} Padoan, P., \& Nordlund, \AA. 2011, ApJ, 730, 40   
\bibitem[Papadopoulos (2010)] {papadopoulos10} Papadopoulos, P. P. \& Pelupessy, F. I. 2010, ApJ, 717, 1037
\bibitem[Piau et al. (2011)] {piau11} Piau, L., Kervella, P., Dib, S., \& Hauschildt, P. 2011, A\&A, 526, 100
\bibitem[Schruba (2011)] {schruba11} Schruba, A., Leroy, A. K., Walter, F. et al. 2011, AJ, 142, 37 
\bibitem[Silk (2009)] {silk09} Silk, J., \& Norman, C. 2009, ApJ, 700, 262
\bibitem[Tabatabaei (2010)] {tabatabei10} Tabatabaei, F. \& Berkhuijsen, E. M. 2010, A\&A, 517,  77 
\bibitem[Tutukov (2006)] {tutukov06} Tutukov, A. V. 2006, ARep, 50, 526
\bibitem[Vink (2001)] {vink01} Vink, J. S., de Koter, A., \& Lamers, H. J. G. L. M. 2001, A\&A, 369, 574 
\bibitem[Vollmer (2011)] {vollmer11} Vollmer, B., \& Leroy, A. K. 2011, AJ, 141, 24 
\bibitem[Wong (2002)] {wong02} Wong, T. \& Blitz, L. 2002, ApJ, 569, 157 

\end{thebibliography}

\end{document}